\documentclass[12pt,a4paper]{article}
\usepackage{amsmath}
\usepackage{graphicx}%
\usepackage{amsfonts}%
\newcommand{\Journal}[5]{#1~#2~\textbf{#3}, #5 (#4)}

\newcommand{\IDD}[1]{\int d^4#1}
\begin{document}
\title{
Forty-fifth anniversary \\ of the Nambu--Jona-Lasinio model}
\author{
M.~K.~Volkov and A.~E.~Radzhabov}
\date{}
\maketitle

\center{\small{
Bogoliubov Laboratory of Theoretical Physics, \\
Joint Institute for Nuclear Research, 141980 Dubna, Russia}}

\abstract{ A short review of the development of the
Nambu--Jona-Lasinio (NJL) model is given. The $SU(2)\times SU(2)$ and
$U(3)\times U(3)$ local quark NJL models are considered. The
mechanisms of spontaneous breaking of chiral symmetry and vector
dominance are shown. The local NJL model allows us to describe the
mass spectrum and main strong and electroweak decays of the four
ground-state nonets of pseudoscalar, scalar, vector and axial-vector
mesons. Applications of this model to the description of mesons in
hot and dense medium are
discussed. It is shown that for solving problems connected with the description of the 
meson radial excitations and quark confinement it is necessary to consider a nonlocal
extension of the NJL model. The main attention is attracted to the description of the methods
used in different versions of the NJL model. Physical results for low-energy hadron physics
obtained in these models can be found in the cited works. }

\section{Introduction}

The NJL model was proposed in 1961, where the authors attempted to explain the origin of the
nucleon mass with the help of spontaneous breaking of chiral symmetry \cite{Nambu}. The model
was formulated in terms of nucleons, pions and scalar $\sigma$ mesons\footnote{It is
worthwhile to note that in the same 1961 two papers devoted to the same problems were
published in the USSR in Moscow: V.G. Vaks and A.I. Larkin \cite{Vaks} and B.A. Arbuzov, A.N.
Tavkhelidze and R.N. Faustov \cite{Arbuzov}.}. Remind that the fundamental theory of strong
interactions, QCD, was not constructed at that time.

After 15 years this model was reformulated at the quark language by the Japanese physicists
T.~Eguchi and K.~Kikkawa \cite{Eguchi,Kikkawa}. It should be noted that it is supposed that
all hadrons are formed from constituent quarks with mass $m \approx 300$ MeV, whereas the QCD
theory is based on more light current quarks with mass $m_0 \approx 5-7$ MeV. In
\cite{Eguchi,Kikkawa}, it is shown that light current quarks turn into the massive constituent
quarks due to spontaneous breaking of chiral symmetry. However, they only considered the quark
NJL model in the chiral limit $m_0=0$. In this limit all masses of pseudoscalar mesons are
equal to zero.

Starting from 1982 M.K.~Volkov and D.~Ebert with collaborators considered a more realistic
version of the quark NJL model when $m_0 \neq 0$
\cite{VolkovYaf82,VolkovZPhys83,VolkovAnnP84}. It allowed them to describe the mass spectrum,
internal properties, and strong and electroweak interactions of scalar, pseudoscalar, and
vector meson nonets \cite{VolkovEch,EberRein}. In 1984, T.~Hatsuda and T.~Kunihiro applied
this model to the description of hadrons in hot dense medium
\cite{Hatsuda:1984jm,Hatsuda1985}.

After 1986 this model gained popularity. In the following years more than six hundred papers
devoted to the NJL model were published. Therefore, we cannot give here a full list of the
corresponding references. We only note some authors who gave a noticeable contribution in
this field of research.

The NJL model was especially intensively used for different
applications in Germany (D.~Ebert, Humboldt Univ., Berlin; H.
Kleinert, Freie Univ., Berlin; H. Reinhardt, Tubingen Univ.; J.
Hufner, S. Klevansky, Heidelberg Univ.; W. Weise Munich, Tech. Univ.;
D. Blaschke, G. Roepke Rostock Univ.; K. Goeke, C.V. Christov Ruhr
Univ., Bochum; B. Friman Darmstadt GSI and others), Japan
(T.~Hatsuda, K.~Yazaki Tokyo Univ.; T.~Kunihiro Kyoto Univ.;
M.~Asakawa, S.~Sawada, K. Yamawaki Nagoya Univ. and others) and USA
(M.D.~Scadron Arizona Univ.; C.M.~Shakin City Coll., N.Y. and
others). The NJL model was also studied by many physicists from
China, England, France, Italy, Portugal, South African Republic and
other countries.

In our country this model was actively developed at JINR Dubna (M.K.
Volkov, G.V. Efimov, M.I. Ivanov, Yu.L. Kalinovsky, A.A. Osipov with
collaborators), St.-Petersburg State Univ. (A.A. Andrianov and V.A.
Andrianov), St.-Petersburg INP (D. Diakonov, V.Yu. Petrov) and
St.-Petersburg Polytechnic Inst. (A.N. Ivanov).

It is worthwhile to note that now NJL is widely used for different applications as in
elementary particle physics as in nuclear physics (see e.g.
\cite{Vogl:1991qt,Gulamov:1993us}).

\section{$SU(2)\times SU(2)$ NJL model}
\subsection{Pseudoscalar and scalar mesons}
Firstly, let us consider the simple $SU(2)\times SU(2)$ NJL model. The corresponding chiral
quark Lagrangian is
\begin{eqnarray}
\mathcal{L}(\bar{q},q)=\bar{q}(x)(i\hat{\partial}_x-m_0)q(x)+\frac{G_\pi}{2}\left((\bar{q}(x)
 q(x))^2+(\bar{q}(x) i \tau_a \gamma^5 q(x))^2 \right), \label{lagr_4q}
\end{eqnarray}
where $\bar{q}(x)=\{u(x),d(x)\}$ are the fields of $u$, $d$ antiquarks, $m_0$ is the current
quark mass, $G_\pi$ is the four-quark coupling constant, $\tau^a$ are the Pauli matrices, and
$\gamma^5$ is the Dirac matrix.

After bosonization the Lagrangian takes the form
\begin{eqnarray}
\mathcal{L}(\bar{q},q,\sigma,\pi)=\bar{q}(x)(i\hat{\partial}_x-m_0+\sigma(x) + i  \gamma_5
\tau_a\pi_a(x) )q(x) -\frac{\sigma^2(x) + \pi_a^2(x) }{2 G_\pi} , \label{fourQLagr}
\end{eqnarray}
where $\sigma(x)$, $\pi_a(x)$ are the scalar and pseudoscalar meson fields. Effective meson
Lagrangian can be obtained from Lagrangian (\ref{fourQLagr}) in one quark-loop approximation.
Here we can see that vacuum expectation value of pion field is equal to zero, whereas
$\sigma(x)$ has a nonzero vacuum expectation value $\langle \sigma \rangle_0=\sigma_0$.
Therefore, it is necessary to redefine this field in order a new physical field
$\sigma^\prime(x)=\sigma(x)-\sigma_0$ have a zero vacuum expectation value $\langle
\sigma^\prime\rangle_0=0$. Excluding linear in $\sigma^\prime(x)$ terms from the Lagrangian we
obtain gap equation
\begin{eqnarray}
\left.\frac{\delta \mathcal{L}}{\delta
\sigma^\prime}\right|_{\sigma^\prime=0}=0,\quad\Rightarrow\quad m_0=m+\sigma_0=m(1- 8 G_\pi
I_1^\Lambda(m) ) \label{gapNJL}
\end{eqnarray}
This equation describes spontaneous breaking of chiral symmetry. As a result, light current
quark mass $m_0$ turns into massive constituent quark mass $m$. Note that all physical
quantities in the NJL model are expressed through the quadratically and logarithmically
divergent integrals $I_1^\Lambda(m) $ and $I_2^\Lambda(m) $
\begin{eqnarray}
I_1^\Lambda(m) &=&\frac{N_c}{(2 \pi)^4} \int \frac{d_E^4 k \Theta(\Lambda^2-k^2)}{m^2+k^2}
=\frac{N_c}{4 \pi^2}\left[\Lambda^2-m^2\ln\left(\frac{\Lambda^2}{m^2}+1\right)\right]\\
I_2^\Lambda(m) &=&\frac{N_c}{(2 \pi)^4} \int \frac{d_E^4 k \Theta(\Lambda^2-k^2)}
{(m^2+k^2)^2}=\frac{N_c}{4
\pi^2}\left[\ln\left(\frac{\Lambda^2}{m^2}+1\right)-\left(1+\frac{m^2}{\Lambda^2}\right)^{-1}\right]
\nonumber
\end{eqnarray}
These integrals are given in Euclidean space; $N_c=3$ is a number of quark colors; $\Lambda$
is the cut-off parameter. It describes the region where spontaneous breaking of chiral
symmetry takes place.

From the Lagrangian (\ref{fourQLagr}) the following expressions for the $\pi$ and $\sigma$
meson masses can be obtained
\begin{eqnarray}
m_\pi^2=g_{\pi qq}^2 \left(\frac{1}{G_\pi}-8I_1^\Lambda(m) \right),\quad
m_\sigma^2=m_\pi^2+4m^2, \label{mpi}
\end{eqnarray}
where $g_{\pi qq}=g_{\sigma qq}=(4 I_2^\Lambda(m) )^{-1/2}$ are the meson renormalization
constants that provide a correct coefficient in the kinetic terms for the meson Lagrangian. It
is easy to see from eqs. (\ref{gapNJL}) and (\ref{mpi})  that the pion mass is proportional to
the first power of current quark mass (Gell-Mann--Oakes--Renner relation). As a result the
pion become the Goldstone particle with zero mass in the chiral limit $m_0=0$.

From the weak pion decay $\pi \to \mu \nu$ the Goldberger-Treiman relation follows:
\begin{eqnarray}
F_\pi=\frac{m}{g_{\pi qq}}\label{Gold_Trei},
\end{eqnarray}
where $F_\pi=93$ MeV is the weak pion decay constant \cite{Eidelman:2004wy}.
\subsection{Vector and axial-vector mesons}
The quark Lagrangian corresponding to the vector and axial-vector mesons is
\begin{eqnarray}
\mathcal{L}(\bar{q},q)=-\frac{G_\rho}{2}\left((\bar{q}(x) \gamma^\mu \tau_a
q(x))^2+(\bar{q}(x) \gamma^\mu \gamma^5 \tau_a  q(x))^2 \right).
\end{eqnarray}
After bosonization this Lagrangian takes the form
\begin{eqnarray}
\mathcal{L}(\bar{q},q,\rho,a_1)=\bar{q}(x)( \gamma^\mu \tau_a \rho^\mu_a(x) + \gamma^\mu
\gamma^5 \tau_a {a_1}^\mu_a)q(x) +\frac{({\rho}^\mu_a) ^2 + ({a_1}^\mu_a) ^2}{2 G_\rho} ,
\end{eqnarray}
where $\rho^\mu_a$, ${a_1}^\mu_a$ are the fields of vector ($\rho$) and axial-vector ($a_1$)
mesons.

Note that in the description of vector and axial-vector mesons a gauge-invariant
regularization must be used \cite{VolkovEch}. The quark loop with two vector vertices defines
the kinetic term of the vector meson and renormalization constant of the vector field $g_{\rho
qq}$. As a result, the simple relation between $g_{\sigma qq}$ and $g_{\rho qq}$ appears
\cite{Kikkawa,VolkovEch}
\begin{eqnarray}
g_{\rho qq}=\sqrt{6}g_{\sigma qq} \label{gr6gs}
\end{eqnarray}
We obtain for the $\rho$-meson mass
\begin{eqnarray}
M_\rho^2=\frac{g_{\rho qq}^2}{4G_\rho}.\label{mrho}
\end{eqnarray}

The renormalization constant of the $a_1$ meson field $g_{a_1qq}$ coincides with $g_{\rho qq}$
and the mass equals
\begin{eqnarray}
M_{a_1}^2=M_\rho^2+6 m^2
\end{eqnarray}

\subsection{$\pi-a_1$ transitions}
In the NJL model there are quark loops with pseudoscalar and axial-vector vertices that
describe $\pi-a_1$ transitions \cite{VolkovEch,Volkov:1988hp,Volkov:1991yb,Bernard:1995hm}.

These transitions lead to nondiagonal terms in meson Lagrangian of the type
$\sqrt{6}m{a_1}^\mu_a(x) \partial^\mu_x \pi_a(x)$. In order to exclude these terms it is
necessary to redefine the axial-vector field as
\begin{eqnarray}
{a_1}^\mu_a(x) = {a_1^\prime}^\mu_a(x) - \frac{\sqrt{6}m}{M_{a_1}^2} \partial^\mu_x \pi_a(x)
\end{eqnarray}
This leads to additional kinetic terms of the pions and to modification of the renormalization
constant $g_{\pi qq}$. Now this constant is not equal to $g_{\sigma qq}$
\begin{eqnarray}
g_{\pi qq}=Z^{1/2} g_{\sigma qq},\quad Z=\left( 1 - \frac{6m^2}{M_{a_1}^2}\right)^{-1}
\label{gpZgs}
\end{eqnarray}
It is interesting to note that allowance for the $\pi-a_1$ transitions does not affect to
Goldberger-Treiman relation.

\subsection{Numerical estimations}
Let us define the model parameters. From eqs. (\ref{Gold_Trei}), (\ref{gr6gs}), (\ref{gpZgs})
we can define constituent quark mass thought the observables $F_\pi=93$ MeV, $g_{\rho
qq}=6.14$ (this value corresponds to the experimental width of $\rho$ meson) and
$M_{a_1}=1.26$ GeV \cite{Eidelman:2004wy}
\begin{eqnarray}
g_{\rho qq}=\sqrt{6}g_{\sigma qq}=\sqrt{\frac{6}{Z}}g_{\pi
qq}=\sqrt{\frac{6}{Z}}\frac{m}{F_\pi}=\sqrt{6\left( 1 -
\frac{6m^2}{M_{a_1}^2}\right)}\frac{m}{F_\pi}.
\end{eqnarray}
Then
\begin{eqnarray}
m^2=\frac{M_{a_1}^2}{12}\left( 1- \sqrt{1-\frac{4g_{\rho qq}^2F_\pi^2}{M_{a_1}^2}}\right)\quad
\Rightarrow \quad m=280\,\mathrm{MeV}
\end{eqnarray}
The parameter $\Lambda$ can be found from the equation
\begin{eqnarray}
g_{\rho qq}=\sqrt{\frac{6}{4I_2}}\quad \Rightarrow \quad\Lambda=1.25\,\mathrm{GeV}
\end{eqnarray}
$G_\pi$ and $G_\rho$ can be found from equations (\ref{mpi}), (\ref{mrho}) for pion and
$\rho$-meson masses. Therefore, $G_\pi=4.9$ GeV$^{-2}$ and $G_\rho=16$ GeV$^{-2}$. The value
of the current quark mass $m_0$ is defined from the gap equation (\ref{gapNJL}), $m_0=3$ MeV.

\section{$U(3)\times U(3)$ NJL model}

In order to introduce strange mesons in the model, it is necessary to replace the Pauli
matrices $\tau_i$ (i=1..3) by the Gell-Mann matrices $\lambda_i$ (i=0..8, where
$\lambda_0=\sqrt{\frac{2}{3}}\mathbf{1}$). Let us remind that the $U_A(1)$ problem exists
connected with a correct description of masses of $\eta$, $\eta^\prime$ mesons. Indeed, by
using the $U(3)\times U(3)$ symmetric Lagrangian we obtain "ideal" singlet-octet mixing for
pseudoscalar isoscalar mesons. Then, one of these states contains only $u$ and $d$ quarks and
the other state contains only the strange quark. This situation contradicts experimental data.

In order to solve this problem it is necessary to add the t`Hooft interaction \cite{'tHooft}
to NJL Lagrangian \cite{Klimt:1989pm,Klev1992,VolkovEch93,EberReinVolk}. As a result the model
for scalar and pseudoscalar meson nonets consist of two Lagrangians
\begin{eqnarray}
\mathcal{L}^{NJL}& =& {\bar q}(i{\hat \partial} - m^0)q + {G\over 2}\sum_{i=0}^8 [({\bar q}
{\lambda}_i q)^2 +({\bar q}i{\gamma}_5{\lambda}_i q)^2], \nonumber\\
\mathcal{L}^{tH}& =&- K \left( {\det}[{\bar q}(1+\gamma_5)q]+{\det}[{\bar q}(1-\gamma_5)q]
\right), \label{Ldet}
\end{eqnarray}
where $\bar{q}=\{\bar{u},\bar{d},\bar{s}\}$ are antiquark fields, $m^0$ is a current quark
mass matrix with diagonal elements $m^0_u$, $m^0_d$, $m^0_s$ $(m^0_u \approx m^0_d)$.

The Lagrangian (\ref{Ldet}) can be rewritten in the form (see \cite{Klev1992})
\begin{eqnarray}
&&\mathcal{L} = {\bar q}(i{\hat \partial} - m^0)q + {\frac{1}{2}} \sum_{i=1}^9[G_i^{(-)}
({\bar q}{\lambda^\prime}_i q)^2 +G_i^{(+)}({\bar q}i{\gamma}_5{\lambda^\prime}_i q)^2] +
\nonumber \\
&&\qquad+ G^{(-)}_{us}({\bar q}{\lambda}_u q)({\bar q}{\lambda}_s q)  + G_{us}^{(+)}({\bar
q}i{\gamma}_5{\lambda}_u q)({\bar q}i {\gamma}_5{\lambda}_s q)~, \label{LGus}
\end{eqnarray}
where
\begin{eqnarray} &&{\lambda^\prime}_i={\lambda}_i ~~~ (i=1,...,7),~~~\lambda^\prime_8 = \lambda_u = ({\sqrt 2}
\lambda_0 + \lambda_8)/{\sqrt 3},\nonumber\\
&&\lambda^\prime_9 = \lambda_s = (-\lambda_0 + {\sqrt 2}\lambda_8)/{\sqrt 3}, \nonumber \\
&&G_1^{(\pm)}=G_2^{(\pm)}=G_3^{(\pm)}= G \pm 4Km_sI_1^\Lambda(m_s), \nonumber \\
&&G_4^{(\pm)}=G_5^{(\pm)}=G_6^{(\pm)}=G_7^{(\pm)}= G \pm 4Km_uI_1^\Lambda(m_u),
\nonumber \\
&&G_u^{(\pm)}= G \mp 4Km_sI_1(m_s), ~~~ G_s^{(\pm)}= G, ~~~ G_{us}^{(\pm)}= \pm 4{\sqrt
2}Km_uI_1^\Lambda(m_u). \label{DefG}
\end{eqnarray}

The t`Hooft interaction leads to the additional terms in the gap equations for the $u$, $s$
quark masses
\begin{eqnarray}
m_u&=&m_u^0 + 8 m_u G I_1^\Lambda(m_u)+32 m_u m_s K I_1^\Lambda(m_u) I_1^\Lambda(m_s)\nonumber\\
m_s&=&m_u^0 + 8 m_s G I_1^\Lambda(m_s)+32  K \left(m_uI_1^\Lambda(m_u)\right)^2
\end{eqnarray}

Now let us first consider  bosonization of the diagonal parts of the Lagrangian (\ref{LGus})
including the isovector and strange mesons. After renormalization of the meson fields we
obtain
\begin{eqnarray}
\mathcal{L}(\pi, K, a_0, K^*_0) &=& -{g_{\pi}^2\over 2G_{\pi}}\sum_{i=1}^3\phi_i - {g_K^2\over
G_K}\sum_{i=4}^7\phi_i - {g_{a_0}^2\over 2G_{a_0}}\sum_{i=1}^3\sigma_i
- {g_{K_0^*}^2\over G_{K_0^*}}\sum_{i=4}^7\sigma_i - \nonumber \\
&&-i{\rm Tr}~\ln\left\{1 -{1\over i{\hat \partial}-M}\left[\sum_{i=1}^7
(g_{\phi_i}i\gamma_5{\lambda_i}{\phi_i} +g_{\sigma_i}{\lambda_i} {\sigma_i}) \right]\right\},
\label{piKa}\\
&&G_{\pi}=G_1^{(+)},\, G_K=G_4^{(+)},\, G_{a_0}=G_1^{(-)},\, G_{K_0^*}=G_4^{(-)},   \nonumber \\
&&g^2_{a_0}=[4I_2^\Lambda(m_u)]^{-1},~~~g^2_{K_0^*}=[4I_2^\Lambda(m_u,m_s)]^{-1}, \nonumber \\
&&I_2^\Lambda(m_u,m_s)= {N_c\over (2\pi)^4}\int d^4_e k {\theta (\Lambda^2 -k^2)
\over (k^2 + m^2_u)(k^2 + m^2_s)}=\nonumber\\
&&={3\over (4\pi)^2(m_s^2-m_u^2)}\left[m_s^2\ln\left({\Lambda^2 \over m_s^2}+1 \right) -
m_u^2\ln\left({\Lambda^2\over m_u^2}+1 \right) \right],
 \nonumber \\
&&g_{\pi}=Z^{1/2}_{\pi}g_{a_0},~~~g_K=Z^{1/2}_Kg_{K_0^*},~~~Z_{\pi}\approx Z_K\approx 1.44
\quad .\nonumber
\end{eqnarray}
where $M$ is a constituent quark mass matrix, $\phi_i$ and $\sigma_i$ are the pseudoscalar and
scalar fields.

As a result, the following expressions for the meson masses are obtained
\begin{eqnarray}
&&M^2_{\pi}=g^2_{\pi}\left[{1\over G_{\pi}} - 8I^\Lambda_1(m_u)\right], \nonumber \\
&&M^2_K=g^2_K\left[{1\over G_K} - 4[I^\Lambda_1(m_u)+I^\Lambda_1(m_s)]\right]+Z(m_s-m_u)^2,
\nonumber \\
&&M^2_{a_0}=g^2_{a_0}\left[{1\over G_{a_0}} - 8I^\Lambda_1(m_u)\right] + 4m^2_u , \\
&&M^2_{K_0^*}=g^2_{K_0^*}\left[{1\over G_{K_0^*}} - 4[I^\Lambda_1(m_u)+
I^\Lambda_1(m_s)]\right]+ (m_u+m_s)^2 . \nonumber \label{Mpi}
\end{eqnarray}

The nondiagonal part of the Lagrangian (\ref{LGus}) has the form
\begin{eqnarray}
{\Delta}\mathcal{L}&&= {1\over 2}\left\{G_u^{(+)}({\bar q}i\gamma_5\lambda_u q)^2 +
2G_{us}^{(+)}({\bar q}i\gamma_5\lambda_u q)({\bar q}i\gamma_5\lambda_s q)+
G_s^{(+)}({\bar q}i\gamma_5\lambda_s q)^2+ \right. \nonumber \\
&&+G_u^{(-)}({\bar q}\lambda_u q)^2 \left. + 2G_{us}^{(-)}({\bar q}\lambda_u q)({\bar
q}\lambda_s q) +
G_s^{(-)}({\bar q}\lambda_s q)^2 \right\} =  \\
&&= ({\bar q}i\gamma_5\lambda_{\alpha} q)T^P_{\alpha\beta} ({\bar q}i\gamma_5\lambda_{\beta}
q)+({\bar q}\lambda_{\alpha} q) T^S_{\alpha\beta}({\bar q}\lambda_{\beta} q),
~~(\alpha=u,s)~~(\beta=u,s), \nonumber \label{Ldelta}
\end{eqnarray}
where
\begin{eqnarray}
&&T^{P(S)}=\frac{1}{2}  \left(
\begin{array}{cc}
G_u^{(\pm)}    & G_{us}^{(\pm)} \\
G_{us}^{(\pm)}  & G_s^{(\pm)}
\end{array}
\right) \label{Tps}.
\end{eqnarray}

After bosonization we obtain
\begin{eqnarray}
&&{\Delta} \mathcal{ L}= -{g_{\eta_{\alpha}}g_{\eta_{\beta}}\over 4}\eta_{\alpha}
(T^P)^{-1}_{\alpha\beta}\eta_{\beta}- {g_{\sigma_{\alpha}}g_{\sigma_{\beta}}\over
4}\sigma_{\alpha}
(T^S)^{-1}_{\alpha\beta}\sigma_{\beta} - \nonumber \\
&&-i~{\rm Tr}\ln \left\{1 + {1\over i{\hat \partial} - M}[g_{\eta_{\alpha}} i\gamma_5
\lambda_{\alpha}\eta_{\alpha} + g_{\sigma_{\alpha}} \lambda_{\alpha}\sigma_{\alpha} ] \right\}
\label{Lbar},
\end{eqnarray}
that leads to the following meson Lagrangian
\begin{eqnarray}
\mathcal{L}&=&-{g_{\eta_{\alpha}}g_{\eta_{\beta}}\over
4}\eta_{\alpha}(T^P)^{-1}_{\alpha\beta}\eta_{\beta}-
{g_{\sigma_{\alpha}}g_{\sigma_{\beta}}\over 4}\sigma_{\alpha}
(T^S)^{-1}_{\alpha\beta}\sigma_{\beta} + \nonumber \\
&&+4I^\Lambda_1(m)(g^2_{\eta_u}\eta_u^2 + g^2_{\sigma_u}\sigma_u^2) +
4I^\Lambda_1(m_s)(g^2_{\eta_s}\eta_s^2 + g^2_{\sigma_s}\sigma_s^2)-\nonumber\\
&&-2(m^2_u\sigma^2_u + m^2_s \sigma^2_s) = -{1\over 2}\left\{
\eta_{\alpha}M^{P}_{\alpha\beta}\eta_{\beta} +\sigma_{\alpha}M^{S}_{\alpha\beta}\sigma_{\beta}
\right\} \nonumber \label{deltaL2} \\
&& M^P_{uu}= g_{\eta_u}^2\left({1\over 2}(T^P)^{-1}_{uu} - 8I^\Lambda_1(m_u) \right),
\nonumber \\
&& M^P_{ss}= g_{\eta_s}^2\left({1\over 2}(T^P)^{-1}_{ss}-8I^\Lambda_1(m_s) \right), \\
&& M^P_{us}= {1\over 2}g_{\eta_u}g_{\eta_s}(T^P)^{-1}_{us}, \nonumber \label{Mpuu} \\
&& M^S_{uu}= g_{\sigma_u}^2\left({1\over 2}(T^S)^{-1}_{uu} - 8I^\Lambda_1(m_u) \right)
+4m^2_u , \nonumber \\
&& M^S_{ss}= g_{\sigma_s}^2\left({1\over 2}(T^S)^{-1}_{ss}-8I^\Lambda_1(m_s)
\right) + 4m^2_s , \label{Msuu}\nonumber \\
&& M^S_{us}= {1\over 2}g_{\sigma_u}g_{\sigma_s}(T^S)^{-1}_{us}. \nonumber\\
&&g_{\sigma_u}=g_{\sigma{\bar q}q},~~~g_{\sigma_s}=[4I_2^{\Lambda}(m_s)]^{-1/2},
~~~g_{\eta_u}=g_{\pi{\bar q}q},~~~g_{\eta_s}=Z^{1/2}g_{\sigma_s}.\nonumber 
\end{eqnarray}
where $\sigma_u$, $\sigma_s$ are the fields of strange and nonstrange isoscalar mesons, and
$\eta_u$, $\eta_s$ are the pseudoscalar fields.

After diagonalization of the Lagrangian (\ref{deltaL2}) we find masses of  the pseudoscalar
and scalar mesons $\eta$, $\eta'$, $\sigma$ and $f_0$
\begin{eqnarray}
M^2_{(\eta,\eta')}={1\over 2}\left[ M^P_{ss} + M^P_{uu} \mp
{\sqrt {(M^P_{ss}-M^P_{uu})^2 + 4(M^P_{us})^2}} \right],   \\
M^2_{(\sigma,f_0)}={1\over 2}\left[ M^S_{ss} + M^S_{uu} \mp {\sqrt {(M^S_{ss}-M^S_{uu})^2 +
4(M^S_{us})^2}} \right]. \label{Meta2}
\end{eqnarray}

Two additional arbitrary parameters appear in the $U(3)\times U(3)$ NJL model: the current
mass of the strange quark $m_0^s$ and the coupling constant $\mathrm{K}$. We can define these
parameters using kaon mass and $\eta$ - $\eta^\prime$ mass difference
\begin{eqnarray}
m_s = 425~{\textrm {MeV}},~~~ K = 13.3~{\textrm {GeV}}^{-5} \label{msK}.
\end{eqnarray}
Parameters $m_u$ , $\Lambda$ and $G_\pi$ remains unchanged, $m_u=280$ MeV, $\Lambda=1.25$ GeV,
$G_\pi=4.9$ GeV$^{-2}$ .

After that we obtain the following estimations for the masses of the pseudoscalar and scalar
mesons
\begin{eqnarray}
&&M_{\pi} = 135~{\rm \textrm{MeV}},~~~M_K = 495~{\rm \textrm{MeV}}, \nonumber \\
&&M_{\eta} = 520~{\rm \textrm{MeV}},~~~M_{\eta'} = 1000~{\rm \textrm{MeV}},\nonumber \\
&&M_{\sigma} = 550~{\rm \textrm{MeV}},~~~M_{f_0} = 1130~{\rm \textrm{MeV}}, \nonumber \\
&&M_{a_0} = 810~{\rm \textrm{MeV}},~~~M_{K_0^*} = 960~{\rm \textrm{MeV}}, \label{sigfaK}
\end{eqnarray}
The experimental data are
\begin{eqnarray}
&&M_{\pi^0}=134.9764\pm0.0006~{\rm \textrm{MeV}},~~~M_{\pi^{\pm}}=139.6~{\rm \textrm{MeV}}, \nonumber \\
&&M_{K^+}=493.677\pm 0.016~{\rm \textrm{MeV}},~~~M_{K^0}=497.672\pm 0.031~{\rm \textrm{MeV}},
\nonumber \\
&&M_{\eta}=547.45\pm 0.19~{\rm \textrm{MeV}},~~~M_{\eta'}=957.77\pm 0.14~{\rm \textrm{MeV}},
\nonumber\\
&&M_{\sigma_0(400-1200)}=400 - 1200~{\rm \textrm{MeV}},~~~M_{f_0(980)}=980\pm 10~{\rm
\textrm{MeV}},
\nonumber \\
&&M_{a_0}=983.5\pm 0.9~{\rm \textrm{MeV}},~~~M_{K_0^*}\sim 800~{\rm \textrm{MeV}}.
\label{MexpfK}
\end{eqnarray}

The model parameters are fixed by the masses of pseudoscalar mesons. At the same time the
scalar meson masses are in a qualitative agreement with experimental data.

Vector and axial-vector mesons in $U(3)\times U(3)$ version of the NJL model can be introduced
as in $SU(2)\times SU(2)$ version. As a result after bosonization of quark Lagrangian we have
for vector meson masses \cite{VolkovEch} 
\begin{eqnarray}
M_\rho^2=\frac{g_\rho^2}{4G_\rho},\,\,\,M_{K^\star}^2=\frac{g_{K^\star}^2}{4G_\rho}+\frac{3}{2}(m_s-m_u)^2
,\,\,\,M_\phi^2=\frac{g_\phi^2}{4G_\rho},
\end{eqnarray}
where $g_\rho=\sqrt{6}g_{a0}$, $g_{K^\star}=\sqrt{6}g_{K^\star_0}$,
$g_{\phi}=\sqrt{6}g_{\sigma_s}$. Note, the quark loops gives contribution to mass only for
$K^\star$ meson. As a result the vector meson masses are in satisfactory agreement with
experimental data
\begin{eqnarray}
M_\rho=770\,\textrm{MeV},\,\,M_{K^\star}=930 \,\textrm{MeV},\,\,M_\phi=1090\,\textrm{MeV}.
\end{eqnarray}

It is worthwhile to note that the general Lagrangian for the strong interactions of four meson
nonets can be expressed in a very compact form
\begin{eqnarray}
\mathcal{L}^{\mathrm{int}}=\frac{1}{4}\mathrm{Tr}\left\{
g^2\left(\left[\left(\bar{\sigma}-\frac{M}{g}\right),\bar{\phi}\right]_-^2-\left[\left(\bar{\sigma}-\frac{M}{g}\right)^2+\bar{\phi}^2\right]^2\right)-\right.\nonumber
\\
-\frac{1}{2}\left(G_V^{\mu\nu}G_V^{\mu\nu}+G_A^{\mu\nu}G_A^{\mu\nu}\right)+
\left[D_\mu\left(\bar{\sigma}-\frac{M}{g}\right)^2+\frac{g_\rho}{2}\left\{\bar{A}_\mu,\bar{\phi}\right\}_+\right]^2\nonumber
\\+
\left.\left[D_\mu \bar{\phi}
-\frac{g_\rho}{2}\left\{\bar{A}_\mu,\left(\bar{\sigma}-\frac{M}{g}\right)\right\}_+ \right]^2
\right\}
\end{eqnarray}
where
\begin{eqnarray}
\bar{a}=\lambda_i a^i,\quad \quad D_\mu a=\partial_\mu a-i\frac{g_\rho}{2}[\bar{V}_\mu,a]\nonumber \\
G_V^{\mu\nu}=\partial_\mu\bar{V}^\nu-\partial_\nu\bar{V}^\mu
 -i\frac{g_\rho}{2}\left(\left[ \bar{V}^\mu,\bar{V}^\nu \right]_- + \left[ \bar{A}^\mu,\bar{A}^\nu
 \right]_-\right),\nonumber \\
G_A^{\mu\nu}=\partial_\mu\bar{A}^\nu-\partial_\nu\bar{A}^\mu
 -i\frac{g_\rho}{2}\left(\left[ \bar{V}^\mu,\bar{A}^\nu \right]_- + \left[ \bar{A}^\mu,\bar{V}^\nu
 \right]_-\right).\nonumber
\end{eqnarray}

The electroweak interactions are introduced in our model in a gauge-invariant manner on the
basis of the original quark Lagrangian (1). This allows us to describe not only the strong
processes (strong decays, $\pi\pi$, $\pi K$ scattering and so on) but also different
electroweak processes such as e.-m. and weak decays, radii, polarizabilities, different rare
processes(for instance $\eta \to \pi^0 \gamma \gamma$).

\section{Vector dominance}

After introducing e.-m. interactions in the NJL Lagrangian the photons can interact with the
charged mesons only through quark loops. In contrast to the mesons, which are composite
objects, the kinetic term for the photons is introduced independently into the Lagrangian (1).
The allowance made for the quark loops leads only to renormalization of the electromagnetic
fields and the charge.

The part of the Lagrangian describing the electromagnetic interactions has the form
\begin{eqnarray}
\mathcal{L}_{em}=-\frac{1}{4} (F_{\mu\,\nu})^2 - i \mathrm{Tr} \ln
\left[1-\frac{e}{i\hat{\partial}-m}Q \hat{A} \right],
\end{eqnarray}
where
\begin{eqnarray}
F_{\mu\,\nu}=\partial_\mu A_\nu-\partial_\nu A_\mu,\label{Ftensor}
\end{eqnarray}
$Q=(\lambda_3+{\lambda_8}/{\sqrt{3}})/2$ is the operator of the quark charge.

\begin{figure}[tb]
\center{\resizebox{0.5\textwidth}{!}{\includegraphics{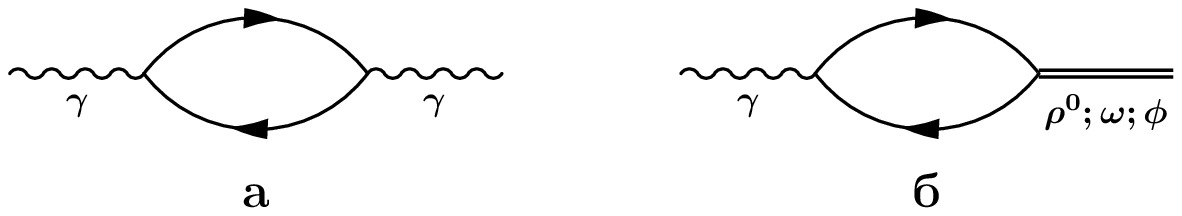}}}
\caption{Divergent quark loops with external photons and vector
mesons $\rho^0$, $\omega$, $\phi$.}
\end{figure}

After calculation of the divergent self-energy diagram of the photon (Fig. 1a), we obtain for
$\mathcal{L}_{em}$ the expression
\begin{eqnarray}
\mathcal{L}_{em}=-\frac{1}{4} (F_{\mu\,\nu}^\prime)^2 - i \mathrm{Tr} \ln
\left[1-\frac{e}{i\hat{\partial}-M}Q \hat{A} \right]^\prime, \label{rirstRen}
\end{eqnarray}
where
\begin{eqnarray}
A_\mu^\prime=\left(1+\frac{4}{3}\frac{e^2}{g_\rho^2}\right)^{1/2} A_\mu^\prime, \quad
e^\prime=\left(1+\frac{4}{3}\frac{e^2}{g_\rho^2}\right)^{-1/2}e
\end{eqnarray}

Besides the self-energy diagrams involving the photon there are divergent diagrams of mixed
type, describing the transitions $\gamma \rho^0$, $\gamma \omega$ and  $\gamma \phi$. (Fig.
1b). The inclusion of these diagrams leads to the appearance in the Lagrangian of terms of the
form
\begin{eqnarray}
\frac{1}{2} \frac{e^\prime}{g_\rho} F_{\mu\,\nu}^\prime
\left(\rho^0_{\mu\,\nu}+\frac{1}{3}\omega_{\mu\,\nu}+\frac{\sqrt{2}}{3}\phi_{\mu\,\nu}\right),
\end{eqnarray}
where $\rho^0_{\mu\,\nu}$, $\omega_{\mu\,\nu}$ and $\phi_{\mu\,\nu}$ are tensors constructed
from meson fields and derivatives, similarly to eq. (\ref{Ftensor}).

As a result, the part of the Lagrangian that describes the electromagnetic interactions of the
mesons and quarks takes the form
\begin{eqnarray}
\mathcal{L}_{em}&=&\frac{M_\rho^2}{2}(\omega_\mu^2+\rho^0_{\mu})^2+\frac{M_\phi^2}{2}\phi_\mu^2
-\frac{1}{4}\left({\rho^0}^2_{\mu\,\nu}+\omega_{\mu\,\nu}^2+\phi_{\mu\,\nu}^2+{F^{\prime}}^2_{\mu\,\nu}\right)\nonumber\\
&&+\frac{1}{2}\frac{e^\prime}{g_\rho}F_{\mu\,\nu}^\prime\left(\rho^0_{\mu\,\nu}+\frac{1}{3}\omega_{\mu\,\nu}+\frac{\sqrt{2}}{3}\phi_{\mu\,\nu}\right)\\
&& -i \mathrm{Tr} \ln \left[1+\frac{1}{i\hat{\partial}-M}
  \left(\frac{g_\rho}{2}(\gamma_\mu\lambda^i V^i_\mu)- e^\prime Q \hat{A}\right) \right]^\prime,\nonumber
\end{eqnarray}
where $V^i_\mu$ are the fields of vector mesons.

We diagonalize the kinetic terms by means of the following substitution of the fields:
\begin{eqnarray}
\rho^0_{\mu}=\tilde{\rho}^0_{\mu}+\frac{e^\prime}{g_\rho}A^\prime_\mu, \quad
\omega_{\mu}=\tilde{\omega}_{\mu}+\frac{e^\prime}{3g_\rho}A^\prime_\mu, \quad
\phi_{\mu}=\tilde{\phi}_{\mu}+\frac{\sqrt{2}e^\prime}{3g_\rho}A^\prime_\mu
\end{eqnarray}
The electromagnetic field and the charge $e^\prime$ are then renormalized as follows:
\begin{eqnarray}
\tilde{A}_\mu&=&\left(1-\frac{4}{3}\frac{{e^\prime}^2}{g_\rho^2}\right)^{1/2} A_\mu^\prime, \label{secondRen}\\
\tilde{e}&=&\left(1-\frac{4}{3}\frac{{e^\prime}^2}{g_\rho^2}\right)^{-1/2}e^\prime
=\left[\left(1-\frac{4}{3}\frac{{e^\prime}^2}{g_\rho^2}\right)\left(1+\frac{4}{3}\frac{e^2}{g_\rho^2}\right)\right]^{-1/2}e
\rightarrow \tilde{e}=e\nonumber
\end{eqnarray}
It is readily seen that as a result of the two renormalizations [ (\ref{rirstRen}) and
(\ref{secondRen}) ] the electric charge takes its original value. The final Lagrangian has the
form
\begin{eqnarray}
\mathcal{L}_{em}&=&\frac{M_\rho^2}{2}(\tilde{\omega}_\mu^2+\tilde{\rho}^0_{\mu})^2+\frac{M_\phi^2}{2}\tilde{\phi}_\mu^2
-\frac{1}{4}\left({{\tilde{\rho^0}}_{\mu\,\nu}}^2+\tilde{\omega}_{\mu\,\nu}^2+\tilde{\phi}_{\mu\,\nu}^2+{{\tilde{F^{\prime}}}}^2_{\mu\,\nu}\right)\nonumber\\
&&+ \left(\frac{e}{3 g_\rho} \right)^2  \left(5 M_\rho^2+m_\phi^2 \right)\tilde{A}^2_\mu
+\frac{e}{g_\rho}\left[M_\rho^2\left(\tilde{\rho^0}_{\mu}+\frac{\tilde{\omega}_{\mu}}{3}\right) +\frac{\sqrt{2}}{3}M_\phi^2\tilde{\phi}_{\mu}\right]\nonumber\\
&&\qquad-i \mathrm{Tr} \ln
\left[1+\frac{1}{i\hat{\partial}-M}\frac{g_\rho}{2}(\gamma_\mu\lambda^i V^i_\mu) \right]
\end{eqnarray}

It is now easy to show that the photons can interact with the charged particles only through
the neutral vector mesons. In this way, we have automatically obtained a model describing
vector dominance. Under the sign of the logarithm, the term with photons has been completely
absorbed by the vector mesons.


\section{Meson in hot and dense matter}

In the last few years the activity in search for a new state of matter, the quark-gluon plasma
(QGP), has significantly increased. New data are already coming from running experiments on
heavy-ion collisions at Brookhaven (RHIC) and CERN SPS. New facilities are planned to be
constructed to increase our capability in this research (LHC, SIS-300). The QGP is expected to
reveal itself through modified properties of hadronic reactions and their products.

The NJL model is a very convenient tool for investigation meson behavior in the hot and dense
matter. First calculations of this type in NJL model were started in
\cite{Hatsuda:1984jm,Hatsuda1985}.

It is possible to use different methods for investigation of meson behavior in the hot and
dense matter. The most popular one is the Matsubara technique \cite{Kapusta}. This "imaginary
time" formalism implies the replacement of the integration over the zero component of the
momentum by the summation of frequencies
\begin{eqnarray}
p^0 &\to& (i \omega_n +\mu)\\
\int \frac{d^4p}{(2 \pi)^4} &\to& i T\sum\limits_n \int \frac{d^3p}{(2 \pi)^3},
\end{eqnarray}
where $\omega_n$ are the Matsubara frequencies, $\omega_n=(2n+1)\pi T$ for fermions and
$\omega_n=2n\pi T$ for bosons; $\mu$ is the chemical potential and $T$ is the temperature.

However, for many applications it is more convenient to use an equivalent representation for
the quark propagator derived in the "real time" formalism \cite{Dolan:1973qd}
\begin{eqnarray}
S(p,T,\mu)&=&(\hat{p}+m)\left[ \frac{1}{p^2-m^2+i\epsilon}+\right.\nonumber\\
&&\left.i 2\pi \delta(p^2-m^2)(\theta(p^0)n(\vec{p},\mu)+\theta(-p^0)n(\vec{p},-\mu))
\right],
\end{eqnarray}
where
\begin{eqnarray}
n(\vec{p},\mu)=\left(1+\exp\frac{E-\mu}{T}\right)^{-1}
\end{eqnarray}
is the Fermi-Dirac function for quarks, $E=\sqrt{\vec{p}^2+m^2}$. This leads to a different
method for calculation of integrals $I_1^{\Lambda_3}(m,T,\mu)$, $I_2^{\Lambda_3}(m,T,\mu)$.
First we performed contour integration in the complex $p_0$ plane and after that reguralized
integrals by three-momentum cut-off $\Lambda_3$. As a result, divergent integrals
$I_1^{\Lambda_3}(m,T,\mu)$ and $I_2^{\Lambda_3}(m,T,\mu)$ take the form
\begin{eqnarray}
I_1^{\Lambda_3}(m,T,\mu)&=& \frac{N_c}{(2\pi)^2}\int \limits_0^{\Lambda_3} dp
\frac{p^2}{E}\left(1-\eta(\vec{p},\mu)-\tilde{\eta}(\vec{p},\mu)\right),\nonumber\\
I_2^{\Lambda_3}(m,T,\mu)&= &\frac{N_c}{2(2\pi)^2}\int \limits_0^{\Lambda_3} dp \frac{p^2}{E^3}\left(1-\eta(\vec{p},\mu)-\tilde{\eta}(\vec{p},\mu)\right).
\end{eqnarray}
We defined values of model parameters in vacuum using the same conditions as in Sect.
2.4\footnote{Note that different methods of regularization give us some other values of model
parameters $m=280$ MeV, $\Lambda_3=1.03$ GeV, $G_\pi=3.48$ GeV$^{-2}$, $G_\rho=15.9$
GeV$^{-2}$ and $m_0=2$ MeV in comparison with sect 2.4.}. Here we assume that model parameters
$G_\pi$, $G_\rho$, $m_0$ and $\Lambda_3$ do not depend on $T$ and $\mu$. The dependence of $m$
on $T$ and $\mu$ is calculated from the gap equation. Then we calculate $T$ and $\mu$
dependence of basic integrals $I_1^{\Lambda_3}(m,T,\mu)$ and $I_2^{\Lambda_3}(m,T,\mu)$. This
allows us to define the dependence on $T$ and $\mu$ of all physical quantities.

\begin{figure}[tb]
\center{\resizebox{0.8\textwidth}{!}{\includegraphics{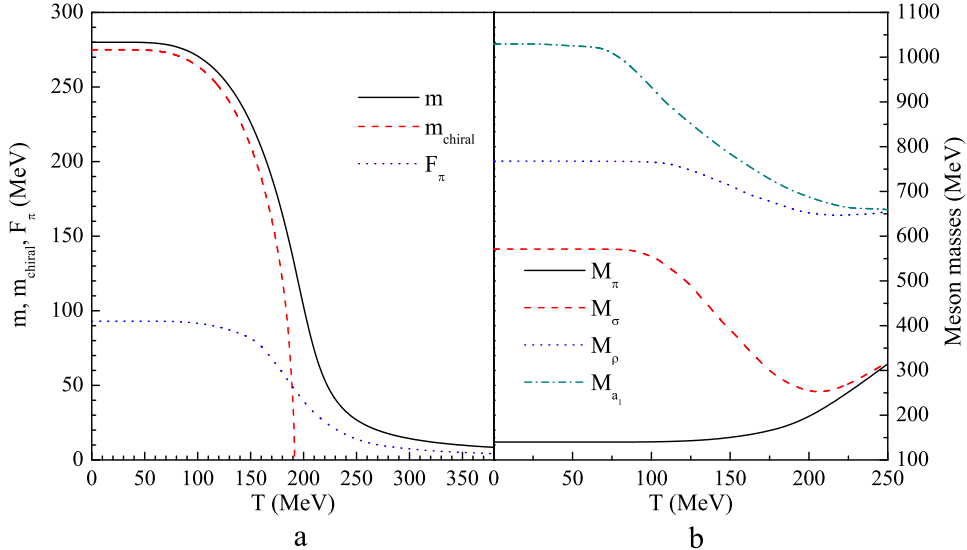}}}
\caption{The behaviour of the quark mass and weak pion decay constant
$F_\pi$ (a), and meson masses $M_{\pi}$, $M_{\sigma}$, $M_{\rho}$,
$M_{a_1}$ (b) as function of $T$. }
\end{figure}

The behaviour of $m(T)$ is shown in Fig.2a. For $m_0 = 0$ the restoration of chiral symmetry
is indicated by the vanishing of the order parameter $m(T)\to 0$ (or quark condensate) at
critical value of the temperature $T_c$. When $m_0 \neq 0$ the sharp phase boundary
disappears. Figure 2b exhibits the behaviour of the meson masses $M_{\pi}$, $M_{\sigma}$,
$M_{\rho}$, $M_{a_1}$ as function of $T$. As $T$ increases the mass of the $\sigma$ meson
decreases sharply with the constituent quark mass. On the other hand, the mass of the pion
will persistently stay constant until the critical conditions for chiral restoration are
reached, beyond which it ceases to exist. $M_\rho$ is merely independent of $T$, whereas
$M_{a_1}$ shows a sharp decrease similar to that of $M_{\sigma}$. Above the critical
temperature one obtains $M_{\pi}=M_{\sigma}$ and $M_{\rho}=M_{a_1}$, as is expected for a
chiral symmetric phase. In \cite{Ebert:1992ag,HatsudaPhysRep} the value for critical
temperature $T_c \approx 200$ MeV. In lattice recent QCD calculations the value for critical
temperature $T_c \approx 170$ MeV \cite{Bernard:2004je}.

Recently, very interesting results have been obtained from the investigation of strongly
interacting quark matter in the color-superconducting phase. We do not consider here the
diquark condensation and the related problems. The reader will find the details in
\cite{Buballa:2003qv,Blaschke:2004cs}.

In conclusion, we would like to note that the properties of some particles can significantly
change when approach to phase transition. The $\sigma$-meson, which in vacuum is a broad
resonance may become very sharp resonance at particular temperature and chemical potential.
When appropriate conditions are reached, this can lead to the amplification of some processes
mediated by the $\sigma$-resonance, such as $\pi \pi \to \gamma \gamma$, $\pi \pi \to \pi \pi$
etc. This amplification, if observed in heavy-ion collision experiment, can be interpreted as
an event indicating on the approaching the QGP.

\section{First radial excitations of mesons}

In the local version of the NJL model, it is impossible to describe radial excitations of
mesons. Therefore, for the description of both the ground and first radially excited states it
is necessary to consider not only the standard local Lagrangian $\mathcal{L}$ (\ref{lagr_4q})
but also additional nonlocal Lagrangian $\mathcal{L}^{\mathrm{nonloc}}$. In Lagrangian
$\mathcal{L}^{\mathrm{nonloc}}$, we introduce a form-factor for each quark-antiquark current
\begin{eqnarray}
J_I(x)=\int \int d^4x_1 d^4x_2 \,  \bar{q}(x_1) \, F_I(x;x_1,x_2) \, q(x_2). \label{J}
\end{eqnarray}
The form-factors $F_I(x;x_1,x_2)$ can be written in a covariant form \cite{Volkov:1996br}.
Here we do not discuss them in details, and only would like to remark the form-factors for the
ground and first radially excited states can be given in a very simple form in momentum space,
\begin{eqnarray}
f_1(k_\bot)&=&\Theta(\Lambda_3-|{k_\bot}|),\label{ffRad}\\
f_2(k_\bot)&=&c(1+d|{k_\bot}|^2)\Theta(\Lambda_3-|{k_\bot}|),\nonumber
\end{eqnarray}
where $k$ is the quark-antiquark pair relative momentum  and $k_{\bot}$ is the part of $k$
transversal to the total momentum $P$
\begin{eqnarray}
k_{\bot}\equiv k- \frac{k \cdot P}{P^2}P
\end{eqnarray}
The step function, $\Theta(\Lambda_3-|{k_\bot}|)$, is a covariant generalization of the
3-dimensional cut-off in the NJL model. For $d<-\Lambda_3^{-2}$ the form-factor $f_2(k_\bot)$
has the form of an excited state wave function with a node in the interval
$0<|{k_\bot}|<\Lambda_3$. The form-factors (\ref{ffRad}) are the first terms in a series
expansion in ${k_\bot}^2$; the inclusion of higher radially excited states would require
polynomials of a higher degree. The factor $c$ describes the change in the strength of the
four-quark interaction in the radially excited channels relative to the constant $G_\pi$. This
constant is defined from the mass of radial pion excitation $M_\pi^\prime = 1300$ MeV. The
parameter $d$ is defined from the condition\footnote{Here, $I_n$, $I_n^f$ and $I_n^{ff}$
denote the loop integrals with zero, one or two factors $f(k_\bot)\equiv f_2(k_\bot)$ in the
numerator
\begin{eqnarray}
I_n^{f..f}\equiv -iN_c \mathrm{tr}\int_{\Lambda_3} \frac{d^4k}{(2\pi)^4}
\frac{f(\mathbf{k})..f(\mathbf{k})}{(m^2-k^2)^n}
\end{eqnarray}
}
\begin{eqnarray}
I_1^f=-iN_c\int_{\Lambda_3} \frac{d^4k}{(2\pi)^4} \frac{f_1(\mathbf{k})}{m^2-k^2}=0
\label{mpi_prime_condition}
\end{eqnarray}
Now let us explain the meaning of this condition. Note that in this model we have two gap
equations
\begin{eqnarray}
\frac{\delta \mathcal{W}}{\delta \sigma_1}=-iN_c \mathrm{tr}\int_{\Lambda_3} \frac{d^4
k}{(2\pi)^4}
\frac{1}{\hat{k}-m_0+\langle\sigma_1\rangle_0+\langle\sigma_2\rangle_0 f_2(\mathbf{k})}-\frac{\langle\sigma_1\rangle_0}{G_\pi}=0 \label{gapRadial}\\
\frac{\delta \mathcal{W}}{\delta \sigma_2}=-iN_c \mathrm{tr}\int_{\Lambda_3} \frac{d^4
k}{(2\pi)^4}
\frac{f_2(\mathbf{k})}{\hat{k}-m_0+\langle\sigma_1\rangle_0+\langle\sigma_2\rangle_0
f_2(\mathbf{k})}-\frac{\langle\sigma_2\rangle_0}{G_\pi}=0\nonumber
\end{eqnarray}
In general, the solution of these equations would have $\langle\sigma_2\rangle_0 \neq 0$; in
this case the dynamically generated quark mass,
$-\langle\sigma_1\rangle_0-\langle\sigma_2\rangle_0 f_1(\mathbf{k})+m_0$, becomes momentum
dependent. Condition (\ref{mpi_prime_condition}) leads to the trivial solution for the second
gap equation and, as a result, we have only one nontrivial gap equation coinciding with the
standard gap equation of the local NJL model.

The free part of the effective action for pions takes the form
\begin{eqnarray}
\mathcal{W}=\frac{1}{2}\int \frac{d^4 P}{(2 \pi)^4}\sum \limits_{i,j=1}^{2}
\pi_i^a(P)K_{ij}^{ab}(P)\pi_j^b(P),
\end{eqnarray}
where $K_{ij}^{ab}(P)\equiv \delta^{ab} K_{ij}(P)$ and
\begin{eqnarray}
K_{11}(P)&=&Z_1(P^2-M_1^2), \quad K_{22}(P)=Z_2(P^2-M_2^2),\\
K_{12}(P)&=&K_{21}(P)=\sqrt{Z_1 Z_2} \Gamma P^2,\nonumber
\end{eqnarray}
with
\begin{eqnarray}
M_1^2&=&\frac{1}{Z_1}\left(\frac{1}{G_\pi}-8I_1\right)=\frac{m_0}{Z_1 G_\pi m},\label{MassM1M2}\\
M_2^2&=&\frac{1}{Z_2}\left(\frac{1}{G_\pi}-8I_1^{ff}\right),\nonumber\\
Z_1&=&4I_2,\,Z_2=4I_2^{ff},\,\Gamma=\frac{4}{\sqrt{Z_1Z_2}}I_2^f.\nonumber
\end{eqnarray}

To determine the physical $\pi$- and $\pi^\prime$-meson states, we have to diagonalize the
quadratic part of the action. It can be performed with the help of an orthogonal tranformation
of the fields $\pi_1$, $\pi_2$; the details of this procedure can be found in
\cite{Volkov:1996br}. We would like to note that after the expansion in a series of a small
current quark mass $m_0$ (from eq. (\ref{MassM1M2}) $M_1^2 \sim m_0$), one finds for the
physical states
\begin{eqnarray}
M_\pi^2&=&M_1^2+O(M_1^4),\\
M_{\pi^\prime}^2&=&\frac{M_2^2}{1-\Gamma^2}\left[1+\Gamma^2\frac{M_1^2}{M_2^2}+O(M_1^4)\right].\nonumber
\end{eqnarray}
Thus, in the chiral limit, the effective Lagrangian indeed describes a massless Goldstone
boson, the pion $\pi$, and a heavy pseudoscalar meson, $\pi^\prime$. The ratio of the $\pi$
and $\pi^\prime$ weak decay  constants can be directly expressed in terms of the physical
meson masses
\begin{eqnarray}
\frac{F_{\pi^\prime}}{F_\pi}
&=&\frac{\Gamma}{\sqrt{1-\Gamma^2}}\frac{M_\pi^2}{M_{\pi^\prime}^2}
\end{eqnarray}
It is worth noting that the matrix elements for a pseudoscalar meson and the divergence of the
axial-vector current must vanish both for the pion and its radial excitation in the chiral
limit. In the case of the pion, the matrix element is $\sim M_\pi^2$ and vanished in the
chiral limit as  $M_\pi^2 \to 0$. The situation is opposite in the case of radially excited
state, there $F_{\pi^\prime} \to 0$ in the chiral limit, while $M_{\pi^\prime}$ remaining
finite.

Here we discuss only pions. In \cite{Volkov:1996fk,Volkov:1999yi,Ebert:2000nx} it is shown
that this method can be extended to the chiral $U(3)\times U(3)$ group for pseudoscalar,
scalar and vector mesons. In the framework of this model the main strong decays of the
radially excited meson were described \cite{Volkov:1997dd,Volkov:1999yi}. One of the most
interesting results obtained in this model concerns the identification of nineteen
experimentally observed scalar states for the masses 0.4 -- 1.7 GeV. These states can be
interpreted in our model as two scalar nonets and a scalar glueball with the mass $\sim 1.5$
GeV \cite{Volkov:2001ns}. The first nonet consists of the ground state mesons with the masses
between 0.4 -- 1 GeV. The second nonet consists of radially excited scalar mesons with the
masses 1.3 -- 1.7 GeV. Four scalar states and scalar glueball are mixed, as they have the same
quantum numbers.

\section{Nonlocal NJL model and quark confinement}

The NJL models have two main defects. They contain ultraviolet (UV)
divergences and do not provide quark confinement. Usually, UV
divergences are removed by using the cut-off parameter $\Lambda$
taken at an energy scale of the order of 1 GeV. The physical meaning
of this cut-off is connected with the separation of the
energy-momentum region, where spontaneous breaking of the chiral
symmetry and bosonization of quarks takes place. In order to exclude
unphysical quark-antiquark thresholds from amplitudes of different
processes only lowest powers of momentum expansion of quark loops are
usually used in the NJL models.

These drawbacks of the standard NJL model can be solved only in the framework of nonlocal
models. There are many different nonlocal versions of NJL model (see e.g.
\cite{Efimov1,Andrianov:1993,Andrianov:1998kj,Celenza:1999cx,Celenza:2000uk}). Here we
demonstrate one version of nonlocal models which is motivated by an instanton interaction
\cite{Radzhabov:2003hy}. Similar models are considered in
\cite{Buballa:1992sz,Plant,DoLT98,Scarpettini:2003fj}.

The $SU(2)\times SU(2)$ symmetric action with the nonlocal four-quark
interaction has the form
\begin{eqnarray}
\mathcal{S}(\bar{q},q)=\int d^4x \,\left\{ \bar{q}(x)(i \hat{\partial}_x -m_c)q(x) +
\frac{G_\pi}{2} \left( J_\sigma(x) J_\sigma(x) + J_\pi^a(x) J_\pi^a(x) \right)\right.-\nonumber \\
\left.-\frac{G_\rho}{2} (J_\rho^{\mu \, a}(x) J_\rho^{\mu \, a}(x) + J_{a_1}^{\mu \, a}(x)
J_{a_1}^{\mu \, a}(x)) \right\}\label{lag},
\end{eqnarray}
The nonlocal quark currents $J_I(x)$ are expressed as
\begin{eqnarray}
J_I(x)=\int \int d^4x_1 d^4x_2 \, f(x_1)f(x_2)\, \bar{q}(x-x_1) \, \Gamma_I \, q(x+x_2) ,
\label{Jnl}
\end{eqnarray}
where the nonlocal function $f(x)$ is normalized by $f(0)=1$.  In (\ref{lag}) the matrices
$\Gamma_I$ are defined as $\Gamma_\sigma=\mathbf{1}$, $\Gamma_\pi^a=i \gamma^5 \tau^a$,
$\Gamma_\rho^{\mu \, a}=\gamma^\mu \tau^a$, $\Gamma_{a_1}^{\mu \, a}=\gamma^5 \gamma^\mu
\tau^a$.

After bosonization the scalar field $\sigma$ will have nonzero vacuum expectation value. In
order to obtain a physical scalar field with zero vacuum expectation value, it is necessary to
shift the scalar field. This leads to the appearance of the nonlocal quark mass $m(p^2)$
instead of the current quark mass $m_0$ which can be found from the gap equation
\begin{eqnarray}
m(p^2)&=&m_0+ G_\pi \frac{2  N_c }{(2 \pi)^4}f^2(p^2) \IDD{k}
\frac{f^2(k^2)m(k^2)}{k^2+m^2(k^2)}
=\nonumber\\
&=& m_c+(m(0)-m_c) f^2(p^2),\label{gap}
\end{eqnarray}
where $m(0)$ is the dimensional parameter which plays the role of the constituent quark mass.
The quark propagator takes the form
\begin{eqnarray}
S(p) = (\hat p - m(p^2))^{-1}. \label{QP}
\end{eqnarray}

We use one of the simplest $\mathrm{ans\ddot{a}tz}$ for the dynamically generated quark
propagator. In the spirit of \cite{Efimov1,Efimov2} we demand that pole singularities are
absent in the vector part of the quark propagator
\begin{eqnarray}
\frac{1}{m^{2}(p^2)+p^{2}}=\frac{1-\exp\left(  -{p^{2}}/{\Lambda^{2}}\right)}{p^2}.
\end{eqnarray}
The expression for $m(p)$ is found to be:
\begin{equation}
m(p)=\left(  \frac{p^{2}}{\exp\left(  {p^{2}}/{\Lambda^{2}}\right)
-1}\right)  ^{1/2}.\label{M(p)}%
\end{equation}
The mass function $m(p^2)$ depends only on one free parameter $\Lambda$, has no any
singularities in the whole real axis and exponentially drops as $p^{2}\rightarrow\infty$ in
the Euclidean domain. From eq.(\ref{gap}) it follows that nonlocal form factors have a similar
behavior providing the absence of UV divergences in the model. At $p^{2}=0$ the mass function
is equal to the cut-off parameter $\Lambda$, $m(0)=\Lambda$. From the gap equation we find the
relation between the four-quark coupling $G_{1}$ and the nonlocality parameter $\Lambda$
\begin{equation}
G_{1}=\frac{2\pi^{2}}{N_{c}}\frac{1}{\Lambda^{2}}.\label{G(L)}%
\end{equation}
Moreover, the expression for pion renormalization constant has a simple form
\begin{align}
g_{\pi}^{-2}(0)=\frac{N_{c}}{4\pi^{2}}\left(\frac{3}{8}+\frac{\zeta
(3)}{2}\right)\label{gpi}%
\end{align}
where $\zeta$ is the Riemann zeta function. In the chiral limit one has two arbitrary
parameters $\Lambda$, $G_{2}$. We fix their values with the help of the weak pion decay
constant $F_{\pi}=93$ MeV, and $\rho$-meson mass $M_{\rho}=770$ MeV. By using the
Goldberger-Treiman relation $g_{\pi }(0)=m(0)/F_{\pi}$ one finds $\Lambda=m(0)=340$ MeV.

This simple model leads to reasonable predictions for the $\sigma$-meson mass $M_\sigma=420$
MeV and strong decay $\rho \to \pi \pi$ $\Gamma_{\rho\pi \pi}= 135 \, \mathrm{MeV}$.

Nonlocal model, in contrast with the local NJL model, can be successfully used for the
description of not only the constant part of amplitudes of meson interactions but also of the
momentum dependence of amplitudes at small energies. It is worth noting that in the nonlocal
model the relative contributions to the pion form-factor $F_{\gamma ^{\ast}\pi^{+}\pi^{-}}$
and pion radii from the contact diagrams and diagrams with vector mesons as intermediate
states have reasonable values \cite{Dorokhov:2003sc}. In the nonlocal model the contribution
of vector mesons is noticeably smaller than that of the contact diagrams in contrast with the
local NJL model where these contributions are comparable \cite{Volkov:1996rc}. The
vector-meson diagrams play a very important role in the description of the pion form-factor
$F_{\gamma^{\ast}\pi^{+}\pi^{-}}$ in the time-like region. These diagrams allow us not only to
describe the $\rho$-meson resonance but also to obtain a correct behavior of the process
form-factor in the region below 1 GeV.

\section{Conclusion}

Once more emphasize that when the first version of the NJL model was proposed, the fundamental
theory of strong interactions QCD did not exist. Therefore, that time different versions of
the phenomenological hadron models were used for description of low-energy hadron physics,
whereas the description of hadron interactions at large energies was very problematic.
However, after the construction of QCD and the discovery of the phenomenon of asymptotic
freedom it became possible to describe an interaction of hadrons at large energies by means of
perturbation theory. The perturbation theory is applicable only for energies larger than $1$
GeV when the strong coupling constant is smaller than unity. Therefore, for description of the
low-energy region the usage of different phenomenological models is again needed.

One of the most attractive models of that kind is the NJL model. The basis of the model is the
chiral symmetry of strong interactions(as in QCD). The region of applicability of this model
supplements the QCD perturbation theory. The joint application of both the QCD theory and the
NJL model allows us to consider the whole energy region of the strong hadron interactions.

M.K. Volkov is grateful to all collaborators (especially to D.~Ebert) who took part in the
construction and development of the NJL model.

The work is supported by RFBR grant 05-02-16699.

\newpage

\newpage

\end{document}